\documentclass[12pt,bibliography=totoc]{scrartcl}

\usepackage{soul}
\usepackage[utf8]{inputenc}
\usepackage[english]{babel}
\usepackage{graphicx,amsmath,amsfonts,amssymb,dcolumn,amsthm,slashed,bbm,xspace,pgffor,tikz,mathbbol,latexsym,cite,braket}
\usepackage{microtype}
\usepackage{mathtools}
\usepackage{nccmath}
\usepackage{lmodern}
\usepackage[normalem]{ulem}
\usepackage{xcolor}
\usepackage[colorlinks=true,citecolor=blue,urlcolor=blue,linkcolor=blue]{hyperref}
\usepackage[hang]{footmisc} 
\usepackage[affil-it]{authblk}
\usepackage{multicol}

\numberwithin{equation}{section}

\begin{document}

\title{\textbf{Hawking-Page transition in holographic QCD at finite density 
}}

\author{Nelson R.~F.~Braga\thanks{\href{mailto:braga@if.ufrj.br}{ braga@if.ufrj.br}},~  Octavio C.~Junqueira\thanks{\href{mailto:octavioj@pos.if.ufrj.br}{octavioj@pos.if.ufrj.br}} }
\affil{ UFRJ --- Universidade Federal do Rio de Janeiro, Instituto de Física,\\
Caixa Postal 68528, Rio de Janeiro, Brasil}

\date{}
\maketitle

\begin{abstract}

We study the confinement/deconfinement transition of QCD matter for a quark-gluon plasma at finite density using AdS/QCD duality. In order to determine the critical temperature and its dependence on the quark chemical potential, the semi-classical Hawking-Page approach is considered for a charged AdS black hole. The result obtained is consistent with the QCD phase diagram, with the critical temperature decreasing as the chemical potential increases. Using the soft wall holographic model, the value of quark density in which the phase transition occurs at zero temperature is estimated. For higher densities the QCD matter is deconfined, independent of the temperature.   
 
\end{abstract}

\section{Introduction}

Heavy-ion collisions, produced in particle accelerators such as the Relativistic Heavy Ion
Collider (RHIC) and the Large Hadron Collider (LHC) led to the discovery of a new state of matter: the quark gluon plasma (QGP). Understanding the factors that affect the confinement/deconfinement transition between the hadronic and quark-gluon plasma (QGP) phases of strongly interacting matter is one of the major goals of high energy physics. In the hadronic phase, the Quantum Chromodynamics (QCD) matter is confined, while in the QGP the quarks and gluons interact strongly but are not confined \cite{Busza:2018rrf}.  The QGP cannot be observed directly. All the information about this special state of matter, that behaves like a perfect fluid, comes from the particles detected after hadronization, see \cite{Bass:1998vz, Shuryak:2008eq, Casalderrey-Solana:2011dxg}.

 Gauge/gravity duality \cite{Maldacena:1997re,Witten:1998qj, Witten:1998zw} motivated the 
 development of many interesting tools for describing some aspects of systems governed by strong interactions, as can be seen, for example, in\cite{ Polchinski:2001tt,Boschi-Filho:2002xih,Boschi-Filho:2002wdj,Sakai:2004cn,Karch:2006pv,Herzog:2006ra,Nakamura:2006xk,Gubser:2008yx,BallonBayona:2007vp, Gursoy:2008bu}. In particular, it is possible to find holographic representations for a plasma with finite density, that corresponds in the QGP to a finite quark chemical potential \cite{Colangelo:2010pe, Colangelo:2011sr,Colangelo:2012jy,Lee:2009bya,Horigome:2006xu,Nakamura:2006xk,Sachan:2011iy,Braga:2017oqw,Fang:2018axm,Zhao:2022uxc, Braga:2019xwl,Ballon-Bayona:2020xls,Mamani:2022qnf}. The gravitational dual representation of QCD matter at finite density is given by a charged anti de-Sitter (AdS) black hole (BH), where the BH charge is related to the quark density in the gauge theory side of the duality. From this type of models, it is possible to investigate the confinement/deconfinement phase transition at large quark density.

In a recent work \cite{Zhao:2022uxc},  the confinement/deconfinement transition in the Einstein-Maxwell-dilaton model was studied, using the dilaton that characterizes the soft wall model\cite{Karch:2006pv}. The critical temperatures at different values of the chemical potential was obtained from the analysis of Polyakov loops. The results obtained are in agreement with the QCD phase diagram, in the sense that the critical temperature decreases as the quark density increases. In reference \cite{Herzog:2006ra}, the critical temperature for a plasma at zero quark density was determined by applying the Hawking-Page transition approach, which is based on the analysis of the regularized black hole free energy.  Our purpose here is to study the confinement/deconfinement phase transition in the soft wall model at finite quark density, following the Hawking-Page approach,  and compare our results with the ones obtained from the Polyakov loop approach in \cite{Zhao:2022uxc}.    

This work is organized as follows: in  Section 2, we present the charged AdS black hole geometry and discuss how the Hawking temperature is affected by the BH charge. In Section 3, we construct the regularized BH action in the soft wall AdS/QCD model at finite density and temperature, in the presence of the heavy vector mesons. The critical temperatures of confinement/deconfinement transition  at different quark chemical potentials are then studied in Section 4, where the results are obtained from numerical analysis. The Hawking-Page curve of the system is obtained. Section 5 contains our final considerations.

\section{Charged AdS black hole geometry}

The geometry that corresponds to a dual description of the quark-gluon plasma with finite density $\mu$ was studied in refs. \cite{Colangelo:2010pe, Colangelo:2011sr,Colangelo:2012jy}, and is given by a 5-D AdS charged black hole space given by the metric
\begin{equation}\label{BHAdS}
ds^2= \frac{L^2}{z^2}\left( -f(z) dt^2 + d\overrightarrow{x}^2 + \frac{dz^2}{f(z)} \right)\;,
\end{equation}
with 
\begin{equation}\label{f(z)}
f(z) = 1 - \frac{z^4}{z_h^4}-q^2 zh^2 z^4 + q^2 z^6\;,
\end{equation}
where $z_h$ is the location of the horizon, such that $f(z_h) = 0$, and $q$ is a parameter that is proportional to the BH charge. This metric is equivalent to the Reissner-Nordstrom (RN) AdS BH, assuming that the BH charge ($Q$) is related to $q$ through, see \cite{Lee:2009bya},
\begin{equation}\label{Q}
Q^2 = \frac{3 g_5^2 L^2}{2 \kappa^2}q^2\;.
\end{equation}

The hadronic (confined) matter phase,  in turn, is described by the thermal AdS space without a black hole, which is given by Eq. \eqref{BHAdS} taking $f(z) = 1$, \textit{i.e.}, by the line element
\begin{equation}\label{thermalAdS}
    ds^2= \frac{L^2}{z^2}\left(-dt^2 + d\overrightarrow{x}^2 + dz^2 \right)\;.
\end{equation} 

Performing the Wick rotation, $t \rightarrow i \tau$, the compactified time  coordinate ($\tau$) is periodic with the BH period given by $\beta = 1/T$, being $T$ the BH temperature, which is associated to the charge $q$ and the horizon position of the black hole according to the Hawking equation
\begin{equation}
T = \frac{\vert f^\prime(z)\vert_{(z = z_h)}}{4\pi} = \frac{1}{\pi z_h} \left(1-\frac{q^2 z_h^6}{2}\right) \;,
\end{equation}
that comes from
the condition of absence of conical singularity at the horizon \cite{Hawking:1982dh}. By requiring that both geometries, the thermal and BH AdS ones, possess the same shape at the asymptotic ultraviolet limit, $z  = \epsilon$ with $\epsilon \rightarrow 0$, one finds that the period of the thermal AdS space  in the Euclidean signature is $\beta^\prime  = \beta \sqrt{f(\epsilon)}$, see \cite{Herzog:2006ra}. With the periods defined in this way, we are able to apply the Hawking-Page approach, and then analyze the confinement/deconfinement phase transition via holography.   

\section{Soft wall AdS/QCD model at finite density}

In the holographic soft wall AdS/QCD model, one introduces an energy parameter that breaks the conformal symmetry of the action. This is done trough the introduction of a dilaton background field, $\Phi(z) = c^2 z^2$, that contains the infrared (IR) energy parameter ($\sqrt{c}$). At finite temperature, the gravity part of the action of the model is written in the general form
\begin{equation}\label{action1}
	I_G = - \frac{1}{ 2 \kappa^2} \int_0^{z_f} dz\int d^4x \sqrt{-g} e^{-\Phi}\left( R - \Lambda \right) \;,    
\end{equation}
where $\kappa$ is the gravitational coupling associated with the Newton constant,
and $\Lambda$, the cosmological constant. The time coordinate runs from $0$ to $\beta$ (or $\beta^\prime$ for the thermal AdS case). The upper limit of the integration over the holographic coordinate ($z_f$) depends on the geometry, such that $z_f = z_h$ for the BH space, while $z_f \rightarrow \infty$ for the thermal AdS one that does not posses an horizon. In AdS spaces, $\Lambda$ and the curvature are related to the AdS radius by 
\begin{equation}\label{R}
    \Lambda = \frac{3}{5} R = -\frac{12}{L^2}\;.  
\end{equation}

The quark chemical potential $\mu $ is defined in such a way that the variation of the energy density of a system with quarks when their density  $ J^0 \equiv \bar{\psi} \gamma^0 \psi$ has an infinitesimal change is   $ \mu \, \delta  J^0  $. So that the effect of finite quark density is reproduced in QCD by adding a term $ \mu \,  J^0  $ to the Lagrangian in the generating functional. In the holographic approach, boundary values of bulk fields act as the sources for the correlators of the associated gauge theory operators.  Since $ \mu $ acts as a source for the quark density, that is the zero component of vector (Abelian) operator, one can introduce an U(1) field in the holographic model such that its boundary value is identified with $\mu$.  
Therefore, in order to represent the quark density in this holographic approach, one introduces an U(1) vector field $V_M = (V_\mu, V_z)$ with $\mu = \{0,1,2,3\}$ and action 

\begin{equation}\label{action2}
    I_{VF} = -\frac{1}{4g_5^2}\int_0^{z_f}dz \int d^4x \sqrt{-g} e^{-\Phi} F_{MN}F^{MN}\;,
\end{equation}
where $F_{MN} = \partial_{M} V_N - \partial_N V_M$, and $g^2_5$ is a five-dimensional gauge coupling constant.

The total action is then
\begin{equation}\label{totalI}
    I = I_G + I_{VF}\;, 
\end{equation}
and one chooses a solution for the vector field in the RN AdS black hole background of the form\cite{Colangelo:2010pe, Lee:2009bya}
\begin{eqnarray}
V_0 &=& A_0(z) \;= i(C - Qz^2)\;,\nonumber \\
V_i &=& V_z = 0 \quad (i = 1,2,3)\;,\label{V} 
\end{eqnarray} 
where $C$ is a constant, and $Q$ is the black hole charge. The $i$-factor appears due to Wick rotation.

As discussed above, in the dual gauge theory the boundary value of the time component of the $U(1)$ gauge field in the bulk, $A_0(z \rightarrow 0)$, is the source for the quark density $\bar{\psi} \gamma^0 \psi$. Then, associating the boundary value of the $A_0 $ field with $\mu$ after the Wick rotation, one has $A_0(0) = i \mu$, from which one finds $C = \mu$. Thus \begin{equation}\label{A0}
    A_0(z) = i(\mu - \eta q z^2)\;,
\end{equation}
where $\eta = \sqrt{\frac{3 g^2_5 L^2}{2 \kappa^2}}$, according to Eq. \eqref{Q}. 

The requirement of a regular norm of the vector field is satisfied imposing the Dirichlet boundary condition $A_0(z_h) = 0$, see \cite{Horigome:2006xu, Nakamura:2006xk, Hawking:1995ap, Ballon-Bayona:2020xls},
which yields the following relation between the charge $q$ and the quark chemical potential:
\begin{eqnarray}\label{muq}
    \mu = \eta q z_h^2 = Q z_h^2\;. 
\end{eqnarray}
This relation will be fundamental to obtain the critical temperatures at different densities of the medium, and to determine the Hawking-Page curve. 

In order to obtain the on-shell action, one must replace Eq. \eqref{R} into Eq. \eqref{action1}, and Equations \eqref{V} and \eqref{A0} into Eq. \eqref{action2}. Having done this, one gets the following total on-shell action for the AdS black hole, 
\begin{eqnarray}\label{IBH}
    I_{BH}(\epsilon) =  V_{3D} \, \beta \int_\epsilon^{z_h}dz \frac{e^{-c z^2}}{z^5}\left( \frac{4L^3}{\kappa^2} + \frac{2L}{g^2_5}Q^2 z^6\right)\;, 
\end{eqnarray}
and, for the thermal AdS space, 
\begin{eqnarray}\label{IADS}
    I_{AdS}(\epsilon) =  V_{3D} \, \beta \sqrt{1 - \frac{\epsilon^4}{z^4} - q^2 z_h^2 \epsilon^4 + q^2\epsilon^6}\int_\epsilon^{\infty}dz \frac{e^{-c z^2}}{z^5}\left( \frac{4L^3}{\kappa^2} + \frac{2L}{g^2_5}Q^2 z^6\right)\;, 
\end{eqnarray}
where $V_{3D}$ is the trivial spatial volume corresponding to coordinates $ \vec x $, and $\epsilon$ is the ultraviolet (Uv) regulator. We need to introduce this regulator because both actions, $I_{BH}$ and $I_{AdS}$, are infinite in the UV limit $\epsilon \rightarrow 0$.

\subsection{Regularized black hole action density}

In order to eliminate the trivial transverse volume, one defines the action density by $\mathcal{E}= I/V_{3D}$, such that 
\begin{eqnarray}
    \mathcal{E}_{BH}(\epsilon) = I_{BH}(\epsilon)/V_{3D} \quad \text{and} \quad \mathcal{E}_{AdS}(\epsilon) = I_{AdS}(\epsilon)/V_{3D}\;, 
\end{eqnarray}
according to Equations \eqref{IBH} and \eqref{IADS}, respectively. Then, to obtain a finite physical quantity, we apply the Hawking-Page approach and define the regularized black hole action density ($\bigtriangleup \mathcal{E}$), given by the difference between the black hole and thermal AdS action densities, namely,
\begin{eqnarray}
    \bigtriangleup \mathcal{E} = \lim_{\epsilon \rightarrow 0}\left[ \mathcal{E}_{BH}(\epsilon) - \mathcal{E}_{AdS}(\epsilon)\right]\;.\label{RegAction}
\end{eqnarray}
This regularized density does not possess UV divergences, which allows us to take the limit $\epsilon \rightarrow 0$. After integrating over the holographic coordinate and using Eq. \eqref{Q}, one finds the expression
\begin{eqnarray}\label{deltaE}
    \bigtriangleup \mathcal{E}(\bar{\beta}, \bar{q},\bar{z_h}) = \frac{e^{-\bar{z_h}^2}\bar{\beta}c^{3/2}L^3}{ \kappa^2  \bar{z_h}^4}\left[(-1+\bar{z_h}^2)-\frac{3}{2} \bar{q}^2 \bar{z_h}^4 + e^{\bar{z_h}^2}\bar{z_h^4}\text{Ei}(-\bar{z_h}^2)\right]\;, 
\end{eqnarray}
wherein we have used the definition of the dimensionless variables:
\begin{eqnarray}
    \bar{z_h} &=& z_h \sqrt{c}\;,\label{var1} \\
    \bar{\beta} &=& \beta\sqrt{c}\;, \\
    \bar{q} &=& q/c^{3/2}\;,\label{var3}  
\end{eqnarray}
and $\text{Ei}(x) = - \int_{-x}^\infty e^{-t}/t\,dt $. If we take the limit of null chemical potential, $\mu \rightarrow 0$ (or $q \rightarrow 0$), we recover the Herzog expression for the regularized BH action density in the soft wall for the case with zero quark density \cite{Herzog:2006ra}, eliminating the contribution that comes from the vector field action.

From the definition of the dimensionless variables \eqref{var1}-\eqref{var3}, we can obtain results that do not depend on the value of the IR energy parameter $\sqrt{c}$. The only condition is that $c$ must be positive, $c >0$, otherwise, the expression of Eq. \eqref{deltaE} is not valid. The stability of the system is determined by the sign of $ \bigtriangleup \mathcal{E}$. When  $ \bigtriangleup \mathcal{E}$ is positive, the free energy associated to the black hole will be greater than the thermal AdS one, meaning that the thermal AdS  configuration is stable. On the other hand, if $ \bigtriangleup \mathcal{E}$  is negative, the black hole is stable. In the gauge/gravity duality, the AdS space with a black role corresponds to the quark-gluon plasma phase, while the thermal AdS, to the confined hadronic one. The confinement/deconfinement phase transition occurs when $ \bigtriangleup \mathcal{E} = 0$. This equation, however, does not possess analytical solution, what compels us to apply numerical methods. 

\section{Critical temperatures and Hawking-Page curve}

The critical temperature depends on the critical horizon position (${z_h}_c$) in which the regularized BH action density vanishes. As we will see, this critical horizon is sensitive to the value of the BH charge (or quark chemical potential).  In order to write $\bigtriangleup \mathcal{E}$ as a function of the dimensionless quark chemical potential
\begin{equation}\label{barmu}
\bar{\mu} = \mu/\sqrt{c} = \eta\bar{q}\bar{z_h}^2
\end{equation}
and $\bar{z_h}$, one must replace 
\begin{eqnarray}\label{barbeta}
    \bar{\beta}  = \frac{\pi \bar{z_h}}{1-\frac{\bar{\mu}^2\bar{z_h}^2}{2}}\;,
\end{eqnarray}
into Eq. \eqref{deltaE}. In this case, one can define
\begin{eqnarray}\label{deltaE2}
    \bar{\bigtriangleup \mathcal{E}}(\bar{\mu},\bar{z_h} ) = \frac{\pi \bar{z_h}}{1-\frac{\bar{\mu}^2\bar{z_h}^2}{2}}H(\bar{\mu},\bar{z_h})\;, 
\end{eqnarray}
where $\bar{\bigtriangleup \mathcal{E}} = \kappa^2/(c^{3/2}L^3) \bigtriangleup\mathcal{E}$, together with
\begin{eqnarray}\label{H}
H(\bar{\mu}, \bar{z_h}) = e^{-\bar{z_h}^2}\left[\frac{(-1+\bar{z_h}^2)}{\bar{z_h}^4}- \frac{3\bar{\mu}^2}{2\eta^2\bar{z_h}^4}  + e^{\bar{z_h}^2}\text{Ei}(-\bar{z_h}^2)\right]\;.
\end{eqnarray}
From now on, we will assume that $\eta =1$. Naturally, the phase transition occurs when
\begin{eqnarray}\label{H1}
    H(\bar{\mu}, \bar{z_h}) = 0 \quad \text{at} \quad \bar{z_h} = \bar{z_h}_c\;,
\end{eqnarray}
being this equation only determined by the BH horizon position and the quark chemical potential. In Figure 1, we plot $H$ as a function of $\bar{z_h}$ at different fixed values of the chemical potential. We observe that the critical horizon positions, the roots of $H(\bar{\mu}, \bar{z_h})$, increases as $\bar{\mu}$ increases.   

\begin{figure}[!htb]
	\centering
	\includegraphics[scale=0.55]{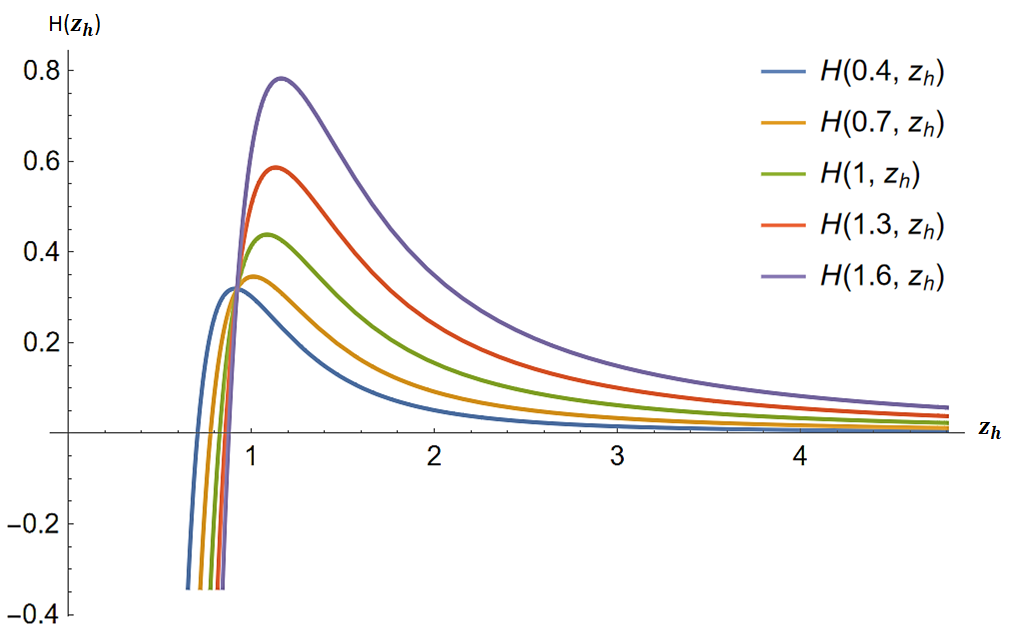}
	\caption{$H(\bar{\mu}, \bar{z_h})$ as a function of $\bar{z_h}$ at different fixed values of the quark chemical potential. The confinement/deconfinement transition occurs at $H(\bar{\mu}, \bar{z_h}) = 0$, \textit{i.e.}, the roots of the respective functions. }
\end{figure}

The values of the critical horizons can be determined by applying numerical methods, and their values are displayed in Table 1 below. From each pair $(\bar{\mu}, \bar{z_c})$, we determine a critical $\bar{\beta}_c$ using Eq. \eqref{barbeta}, and the corresponding dimensionless critical temperature is then computed through the relation \begin{equation}\label{barTc}
\bar{T}_c = T_c/\sqrt{c} = 1/\bar{\beta}_c\;. 
\end{equation} 
This result shows that $T_c$ decreases as the quark chemical potential increases. Such a description is in agreement with the QCD phase diagram. As it is well known, as the density of the medium increases, the temperature where the transition between the hadronic and plasma phases occurs decreases.    

In Figure 2, we plot the points that describe the behaviour of the Hawking-Page curve, using the data from Table 1. The region below the curve corresponds to the confined phase of QCD matter, while the region above represents the quark-gluon plasma.

To obtain the physical values of $T_c$ and $\mu$, we must determine the value of $\sqrt{c}$ using QCD phenomenology. For the soft wall model describing light mesons, this has already been done in \cite{Herzog:2006ra}. The IR energy parameter can be fixed at zero density, from the mass of the lightest $\rho$-meson, that yields $\sqrt{c}=338\, \text{MeV}$. The physical temperatures are then evaluated  multiplying $\bar{T}_c$ by this value. At zero density, one finds $T_c(\mu = 0) = 0.491728 \sqrt{c} = 191\, \text{MeV}$, see \cite{Herzog:2006ra}. This prediction is in accordance with lattice predictions \cite{Karsch:2006xs}.

\begin{table}[h]
\centering
\begin{tabular}[c]{|c|c|c|}
\hline 
  $\bar{\mu}$ & $\bar{z_h}_c$ & $\bar{T}_c$    \\
 \hline
 $\,\,\,0.000000\,\,\, $ &$\,\,\,0.647329\,\,\,$ & $\,\,\, 0.491728\,\,\, $  \\
\hline
 $\,\,\,0.081649\,\,\,$ & $\,\,\,0.651867\,\,\,$ & $\,\,\,0.487613\,\,\, $  \\
\hline
 $\,\,\,0.163299\,\,\,$ & $\,\,\,0.664767\,\,\,$ & $\,\,\,0.476008\,\,\, $  \\
\hline 
 $\,\,\,0.244948\,\,\,$ & $\,\,\,0.684109\,\,\,$ & $  0.458759 $    \\
\hline
 $\,\,\,0.326598\,\,\, $ & $\,\,\,0.707323\,\,\,$ & $0.438013$   \\
\hline
 $\,\,\,0.408248\,\,\,$ &$\,\,\,0.731865\,\,\,$ &$ 0.415517 $  \\ 
\hline
 $\,\,\,0.489897\,\,\,$ &$\,\,\,0.755725\,\,\,$ &$ 0.392332 $    \\
\hline 
$\,\,\,0.571547\,\,\, $ &$\,\,\,0.777637\,\,\,$ &$  0.368900 $  \\
\hline
$\,\,\,0.653197\,\,\,$ &$\,\,\, 0.797013\,\,\,$ &$ 0.345257 $ \\
\hline
 $\,\,\,0.734846\,\,\,$ &$\,\,\,0.813743\,\,\,$ & $0.321232 $  \\ 
\hline
$\,\,\,0.816497\,\,\,$ &$\,\,\,0.827991\,\,\,$ & $ 0.296584 $   \\
\hline 
 $\,\,\,0.898146\,\,\,$&$\,\,\,0.840045\,\,\,$ & $0.271071$   \\
\hline
 $\,\,\,0.979795\,\,\,$ &$\,\,\,0.850223\,\,\,$ &$ 0.244480 $  \\ 
\hline
 $\,\,\,1.061445\,\,\,$ &$\,\,\,0.858825\,\,\,$ &$ 0.216634 $    \\
\hline 
 $\,\,\,1.143095\,\,\,$ &$\,\,\,0.866117\,\,\,$ &$  0.187394 $  \\
\hline
 $\,\,\,1.224744\,\,\,$ &$\,\,\, 0.872325\,\,\,$ &$0.156646$ \\
\hline
$\,\,\,1.306394\,\,\,$ &$\,\,\,0.877634\,\,\,$ & $0.124304 $  \\ 
\hline
 $\,\,\,1.388044\,\,\,$& $\,\,\,0.878122\,\,\,$ & $ 0.102522 $   \\
\hline 
 $\,\,\,1.469693\,\,\,$& $\,\,\,0.870546\,\,\,$ & $ 0.066372 $   \\
\hline 
 $\,\,\,1.551343\,\,\,$& $\,\,\,0.875152\,\,\,$ & $ 0.028507 $   \\
\hline 
 $\,\,\,1.610000\,\,\,$& $\,\,\,0.878122\,\,\,$ & $ 0.000224 $   \\
\hline 
\end{tabular}   
\caption{Quark chemical potentials, critical horizon positions, and the corresponding critical temperatures of confinement/deconfinement phase transition: dimensionless variables.}
\label{table1}
\end{table} 

An interesting point to investigate is the prediction for phase transitions that occur at high densities and very low temperatures. Performing a numerical analysis one obtains that the phase transition at zero temperature occurs between $\bar{\mu} = 1.6102$ and $\bar{\mu} = 1.6104$. Within this precision, we estimate that the transition at zero temperature occurs at

\begin{eqnarray}
   \bar{\mu}_0 \equiv \bar{\mu}(T=0) = 1.610\;, 
\end{eqnarray}
that is, 
\begin{eqnarray}
     \mu_0 = 1.610\sqrt{c} = 544.18\, \text{MeV}\;. \label{mu0}
\end{eqnarray}
This is the holographic prediction, based on the Hawking-Page method, for the quark density in which the transition between the confined hadronic and deconfined plasma phases occurs at approximately zero temperature. For higher values of $\mu $ the QCD matter is a  quark-gluon plasma. 

The relation between the quark and baryon densities is $\mu = \mu_B/3$, see \cite{Colangelo:2010pe}. Therefore, we estimate that the transition at zero temperature occurs at baryon density $\mu_B \approx 1632,54 \, \text{MeV}$. From Polyakov loops in the $2+1$ flavor model \cite{Zhao:2022uxc}, it is predicted that the transition without plasma rotation would occur approximately at $\mu_B \approx 1200 \, \text{MeV}$, see Figure 6 of \cite{Zhao:2022uxc}. Our prediction is valid assuming that $\eta=1$ in the vector field solution \eqref{A0}. Other predictions could be obtained by varying the parameter $\eta$, that changes the relation between $q$ and the chemical potential. The fact that $\eta$ affects the confinement/deconfinement transition was observed in \cite{Colangelo:2010pe}. The predictions would have agreed perfectly if we had assumed $\eta \approx 0.7351$. We must observe that the action used in \cite{Zhao:2022uxc} is not the same as ours, since they have considered an AdS/QCD model for a Einstein-Maxwell-Dilaton gravitational system, see Eq. (2.1) of \cite{Zhao:2022uxc}. The interesting part of the result is that both holographic models predict a baryon density at the same energy scale.  

\begin{figure}[!htb]
	\centering
	\includegraphics[scale=0.75]{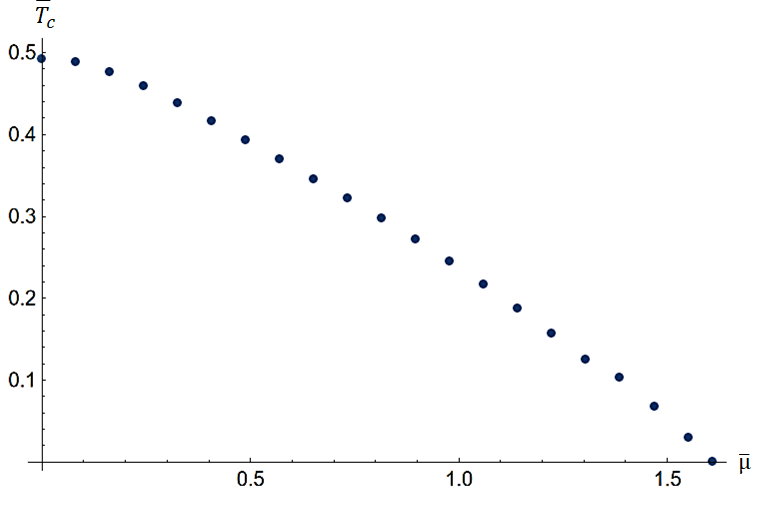}
	\caption{Hawking-Page transition: Critical temperature versus quark density in the soft wall AdS/QCD model. }
\end{figure}

It is important to make it more clear what was the methodology used in this work in order to find the dependence of the critical temperature on the chemical potential. The results were obtained by using the holographic soft wall model for a plasma at finite temperature and chemical potential. The chemical potential is represented by the boundary value of a U(1) gauge field living in a Reissner-Nordstrom space, that corresponds to a charged black hole. The transition from the deconfined phase to the confined one corresponds, in the holographic approach, to a Hawking-Page transition from the BH geometry to a thermal AdS one.  In Eq. \eqref{RegAction} a regularized action density $\bigtriangleup \mathcal{E}$ was defined as the difference between the BH and the the thermal AdS  action densities. So, when the actions are equal this expression vanishes. Therefore the critical temperature was found by searching for the combinations of values of $T$ and $\mu $ that lead the vanishing of the regularized action density. 

The computation of the roots of $\bigtriangleup \mathcal{E}$, the  regularized action density, see Eq. \eqref{deltaE2}, could not be computed analytically. For each fixed value of the BH charge (or the chemical potential), the corresponding critical horizon was obtained by applying numerical methods to Eq. \eqref{H1}. Then the corresponding critical temperatures $\bar{T}_c$ were obtained by using  Eq. \eqref{barbeta}. Using this method one obtains Figure 2 that represents the pairs $(T_c,\, \mu)$, where the values of $\mu$ were previously fixed varying the dimensionless BH charge by $\delta \bar{q} = 0.1$ between each point, with $\bar{\mu}$ defined by Eq. \eqref{barmu} at the corresponding critical horizon. The physical values of $T_c$ and $\mu$ were obtained after fixing the infrared parameter $\sqrt{c}$, which can be done using QCD phenomenology. As mentioned before, here we used the estimate of Ref. \cite{Herzog:2006ra}. An important finding of the present work is the critical quark chemical potential $\mu_0$ of Eq. \eqref{mu0} where the transition occurs at zero temperature. As explained above, this result could be made equal to the one obtained in Ref. \cite{Zhao:2022uxc} by fine tuning the parameter $\eta$ in the vector field solution \eqref{A0}. 

\section{Final considerations}

Holographic models for confinement/deconfinement phase transition in high-energy physics at finite density were studied in \cite{Colangelo:2012jy,Lee:2009bya,Horigome:2006xu,Fang:2018axm,Zhao:2022uxc, Sachan:2011iy}. The authors of \cite{Colangelo:2012jy} analyzed the scalar glueball and vector meson spectral functions in the soft wall model. The confinement/deconfinement transition in the hard wall model was investigated in \cite{Lee:2009bya,Horigome:2006xu}, by introducing a IR cutoff in the holographic coordinate direction. In both cases, the  authors considered the RN/AdS metric solution, that was also used here, see Equations \eqref{V}-\eqref{A0}. In \cite{Sachan:2011iy}, the geometry of a charged black hole is not considered, and a prediction of the quark density at low critical temperatures is not presented. The chiral phase transition in an improved soft wall was investigated in \cite{Fang:2018axm}, for a model with $2+1$ flavors, in which the authors also studied the confinement/deconfinement transition, but not considering the Hawking Page transition between the black hole geometry and the thermal AdS.  A similar analysis was performed in \cite{Zhao:2022uxc}, in which the phase transition was investigated through Polyakov loops, taking into account rotation effects. Again, the critical temperatures at different densities showed to be in harmony with QCD phase diagram, with $T_c$ decreasing as $\mu$ increases. 

In the present work, we have applied the the Hawking-Page approach to the soft wall model in the case of finite quark density, in a similar way as it was was done for the zero density case in \cite{Herzog:2006ra}. We construct, a regularized black hole action density, from which the UV divergencies are eliminated, allowing us to apply numerical methods to study the plasma thermodynamics at very low critical temperatures and large densities. Such a construction is possible in AdS/QCD setup through the geometry of a charged AdS black hole in the gravitation side of the duality, see Eq. \eqref{BHAdS}, being the BH charge associated to the quark chemical potential, according to the relation \eqref{muq}. Differently of \cite{Herzog:2006ra}, where the author assumed that $\mu = 0$, we have considered the contribution of the vector mesons to the total action, as discussed in Section 3. 

The total regularized action density is given by Eq. \eqref{deltaE}, that defines the critical temperatures at different densities, when the regularized free energy density of the charged BH vanishes. As the quark density increases, we observe that the horizon positions at the phase transition also increases, see Table 1. By this reason, the critical temperatures decrease as $\mu$ increases, being this result in harmony with QCD phase diagram, see Hawking-Page curve of Figure 2. This shows the power of the Hawking-Page method, based on the holographic description of dense matter, in agreement with the results obtained in  \cite{Zhao:2022uxc} from the analysis of Polyakov loops.  

Throughout the paper, we have worked with dimensionless variables, as defined by Equations \eqref{var1}-\eqref{var3}, \eqref{barmu} and \eqref{barTc}. In this way, our results are independently of the IR parameter introduced by the dilaton background field, under only the condition that it must be positive. This parameter can be fixed using QCD phenomenology. For the soft wall model, it is estimated that $\sqrt{c}=338\, \text{MeV}$, thus determined by the mass of the lightest $\rho$-meson, see \cite{Herzog:2006ra}. This value leads to a critical temperature at $\mu = 0$ in complete agreement with lattice QCD data, see \cite{Herzog:2006ra, Karsch:2006xs}. This matching brings confidence for the new holographic predictions of critical temperatures of confinement/deconfinement transition at finite quark density, displayed on Table 1, where the same approach of \cite{Herzog:2006ra} was applied. As the quark density increases, there is an interesting point where the phase transition occurs at zero temperature. From a numerical analysis in the soft wall model, we estimate that it happens at quark density $\mu_0 = 544.18\, \text{MeV} $. Our result was obtained assuming that $\eta$ (the free parameter that appears in the RN solution for the vector meson field) is equal to one, see Eq. \eqref{A0}. Then, for $\mu \geq \mu_0$, the QCD matter will be described by a quark-gluon plasma, even at very low temperatures. The present holographic prediction of $\mu_0$ is at the same energy scale of the one obtained via Polyakov loops in \cite{Zhao:2022uxc}, where the authors have used an EMD gravitational model. 

Finally, there is a very important point that deserves a discussion. We have assumed in this work that the transition from a thermal anti-de Sitter (AdS) geometry to a black hole geometry corresponds, in the dual gauge theory, to a transition from a confined to a deconfined phase. This idea appeared first in \cite{Witten:1998zw}, by considering the behaviour of Polyakov loops, that correspond to temporal Wilson loops, in the two geometries. For a pure gauge theory with compactified Euclidean time,  confinement is associated with the vanishing of the Polyakov loops, that is equivalent to the preservation of invariance under the center symmetries. Preservation of such symmetries requires vanishing of the loop, that  corresponds to associating an infinite free energy to the process of adding a color charge to the gauge theory. This represents confinement. Witten realised that the Polyakov loops vanish for the thermal AdS geometry but not for the black hole case \cite{Witten:1998zw}. It is very important to remark that the duality considered by Witten involves a $SU(N)$ gauge theory in the large $N$ limit. The possibility of interpreting the transition between the two geometries as representing a confinement/deconfinement transition for the QCD case, where $N = 3$, is actually an open problem. 

Another important point to be remarked is that it is known that  dynamical fermions explicitly break the center symmetry, because they get a minus sign when a translation of a period  $\beta$ is performed in the time variable. So, including a quark density in the medium would, in principle, be a problem for the interpretation used in this work. 
The quark chemical potential is introduced as the dual of the time component of a quark  current. We considered that this current can be taken as some external background, with no dynamics in the holographic model. This way it was assumed in this work that the interpretation of the thermal AdS and black hole geometries  as dual to the confined and deconfined phases of the gauge theory still holds in this finite density case. The validity of this assumption in real QCD with fermions deserves future discussion. For interesting discussions on this point see, for example,  \cite{Aharony:1998qu,DElia:2005sfk,Conradi:2007kr,Cohen:2014swa}.

\noindent \textbf{Acknowledgments}: The authors are supported by FAPERJ --- Fundação Carlos Chagas Filho de Amparo à Pesquisa do Estado do Rio de Janeiro, CNPq - Conselho Nacional de Desenvolvimento Cient\'ifico e Tecnol\'ogico. This work received also support from  Coordena\c c\~ao de Aperfei\c coamento de Pessoal de N\'ivel Superior - Brasil (CAPES) - Finance Code 001.

\end{document}